\documentclass[preprint,showpacs,preprintnumbers,amsmath,amssymb,natbib]{revtex4}

\usepackage{graphicx}
\usepackage{epsfig}		
\usepackage{dcolumn}
\usepackage{bm}
\usepackage{amsmath}

\def\0{\mbox{\tiny $0$}}
\def\1{\mbox{\tiny $1$}}
\def\2{\mbox{\tiny $2$}}
\def\3{\mbox{\tiny $3$}}
\def\4{\mbox{\tiny $4$}}
\def\5{\mbox{\tiny $5$}}
\def\6{\mbox{\tiny $6$}}
\def\7{\mbox{\tiny $7$}}
\def\8{\mbox{\tiny $8$}}
\def\9{\mbox{\tiny $9$}}

\def\f14{\mbox{\tiny $\frac{1}{4}$}}

\begin{document}

\title{Schr\"odinger cat and Werner state disentanglement simulated by trapped ion systems}

\author{Victor A. S. V. Bittencourt}
\email{vbittencourt@df.ufscar.br}
\author{Alex E. Bernardini}
\email{alexeb@ufscar.br}
\affiliation{Departamento de F\'{\i}sica, Universidade Federal de S\~ao Carlos, PO Box 676, 13565-905, S\~ao Carlos, SP, Brasil}

\begin{abstract}
Disentanglement and loss of quantum correlations due to one global collective noise effect are described for two-{\em qubit} Schr\"odinger cat and Werner states of a four level trapped ion quantum system.
Once the Jaynes-Cummings ionic interactions are mapped onto a Dirac spinor structure,
the elementary tools for computing quantum correlations of two-{\em qubit} ionic states are provided.
With two-{\em qubit} quantum numbers related to the total angular momentum and to its projection onto the direction of an external magnetic field (which lifts the degeneracy of the ion's internal levels), a complete analytical profile of entanglement for the Schr\"odinger cat and Werner states is obtained.
Under vacuum noise (during spontaneous emission), the two-{\em qubit} entanglement in the
Schr\"odinger cat states is shown to vanish asymptotically.
Otherwise, the robustness of Werner states is concomitantly identified, with the entanglement content recovered by their noiseless-like evolution.
Most importantly, our results point to a firstly reported sudden transition between classical and quantum decay regimes driven by a classical collective noise on the Schr\"odinger cat states, which has been quantified by the geometric discord.
\end{abstract}

\date{\today}
\maketitle

\section{Introduction}

Entanglement and coherence are two essential quantum features that support the comprehension of the interface between classical and quantum worlds \cite{001, 002, 003, 004}.
They are widely considered in the proposals for engendering quantum information protocols related to quantum cryptography \cite{A002,A002B}, quantum teleportation and quantum computing processes \cite{A004}.
Since the entanglement brings a suitable kind of overspread non-local coherence, it can intrinsically, and indirectly, mitigate the realization of quantum computing/information processing.

From the experimental perspective, the entanglement and coherence quantifying variables have already been demonstrated to be testable at trapped ion platforms adapted for detecting local quantum correlations \cite{Nat01,Nat03}, and for simulating open quantum systems and quantum phase transitions \cite{Nat02,Nat04,Nat05}.
The ion-trap technology has indeed provided a phenomenological access to manipulate quantum correlational properties of trapped ions \cite{n004,n005,n006,nossopaper}.
Once mapped onto the structure of the Dirac equation, the ionic (anti)Jaynes-Cummings ((A)JC) Hamiltonian dynamics simulates a series of relativistic-like quantum effects \cite{n001,n002,new01}.
Conversely, the corresponding Dirac-like $\mbox{SU}(2)\otimes \mbox{SU}(2)$ bi-spinor structure exhibited by ionic systems encodes the entanglement of two-{\em qubit} states with quantum numbers related to the total angular momentum and to its projection onto the direction of a magnetic field applied to lift the ion's internal levels \cite{nossopaper,n009,n010}.

Trapped ions are not a perfect closed system. Interactions with the environment, which can arise for instance due to random fluctuation of the lifting magnetic field, generate decoherence and degradation of quantum correlations between the subsystems \cite{intronoise00, intronoise01, intronoise02}. From both theoretical and applied perspectives, trapped ion systems were devised to avoid decoherence and dephasing generated by a global noise in setups such as quantum frequency estimation \cite{FreqEst}, for engineering a qubit memory \cite{NoiseTrap01, NoiseTrap02, NoiseTrap03, NoiseTrap04} and the creation and manipulation of a $14$ qubit setup \cite{NoiseTrap02}. Therefore, to clarify how the local decoherence is related to a non-local disentanglement of the trapped ion two-{\em qubit} structure, one is inclined to consider the influence a global collective noise dynamics in more realistic calculations \cite{Yu01}.

Along this paper, the informational content of Schr\"odinger cat and Werner states of a four level trapped ion shall be investigated.
Asymptotic disentanglement due to such a non-local decoherence effect can be identified and estimated through auxiliary quantum correlation quantifiers.
The noise model is evaluated via Kraus operators and the complete dynamics of the trapped ion in the Schr\"odinger representation is obtained through the connection between the (A)JC dynamics and the Dirac equation structure.
The reconstructed Dirac-like spinor density matrix provides the elements for obtaining the quantum entanglement quantified through negativity \cite{n019,QC01} and, when convenient, through geometric discord \cite{QC04}.
On one hand, the Werner states are shown to be unaffected by the collective noisy channel and the profile of quantum oscillations related to entanglement and survivor probabilities are recovered by an equivalent noiseless evolution.
On the other hand, the Schr\"odinger cat states asymptotically disentangle, and the geometric discord indicates a sudden transition between classical and quantum decay regimes.
For the noisy channel here considered, this phenomenon has been reported for the very first time.

The paper is organized as follows. In Section II the correspondence between the Jaynes-Cummings Hamiltonian and the Dirac Hamiltonian with external potentials is identified. Once the map between both dynamics is established, the construction of Dirac Hamiltonian eigenstates allows one to describe the evolution of any initial ionic state, as well as the evaluation of any observable. The main results are presented along Section III, where the global noise, generated by a random, stochastic magnetic field, is included in the ionic dynamics by means of Kraus operator formalism. The survivor probabilities for initial Werner and Cat state are computed, the dynamical evolution of the entanglement and of quantum correlations are recovered by means of, respectively, the negativity and the geometric discord. Our final conclusions are drawn in Section IV.

\section{The map between Jaynes-Cummings Hamiltonian and the Dirac Hamilonian with external potentials}

Relativisc quantum dynamics described by Dirac Equation can be simulated by a four level trapped ion. In this framework, the trapped ion dynamics is described by a total Hamiltonian $\hat{H}_{RJC}$, which includes three interactions between the vibrational degrees of freedom and the internal levels of the ion, is mapped into a Dirac-like Hamiltonian including external potentials.

\begin{figure}[h]
\vspace{-.4cm}
\includegraphics[width = 5.5 cm]{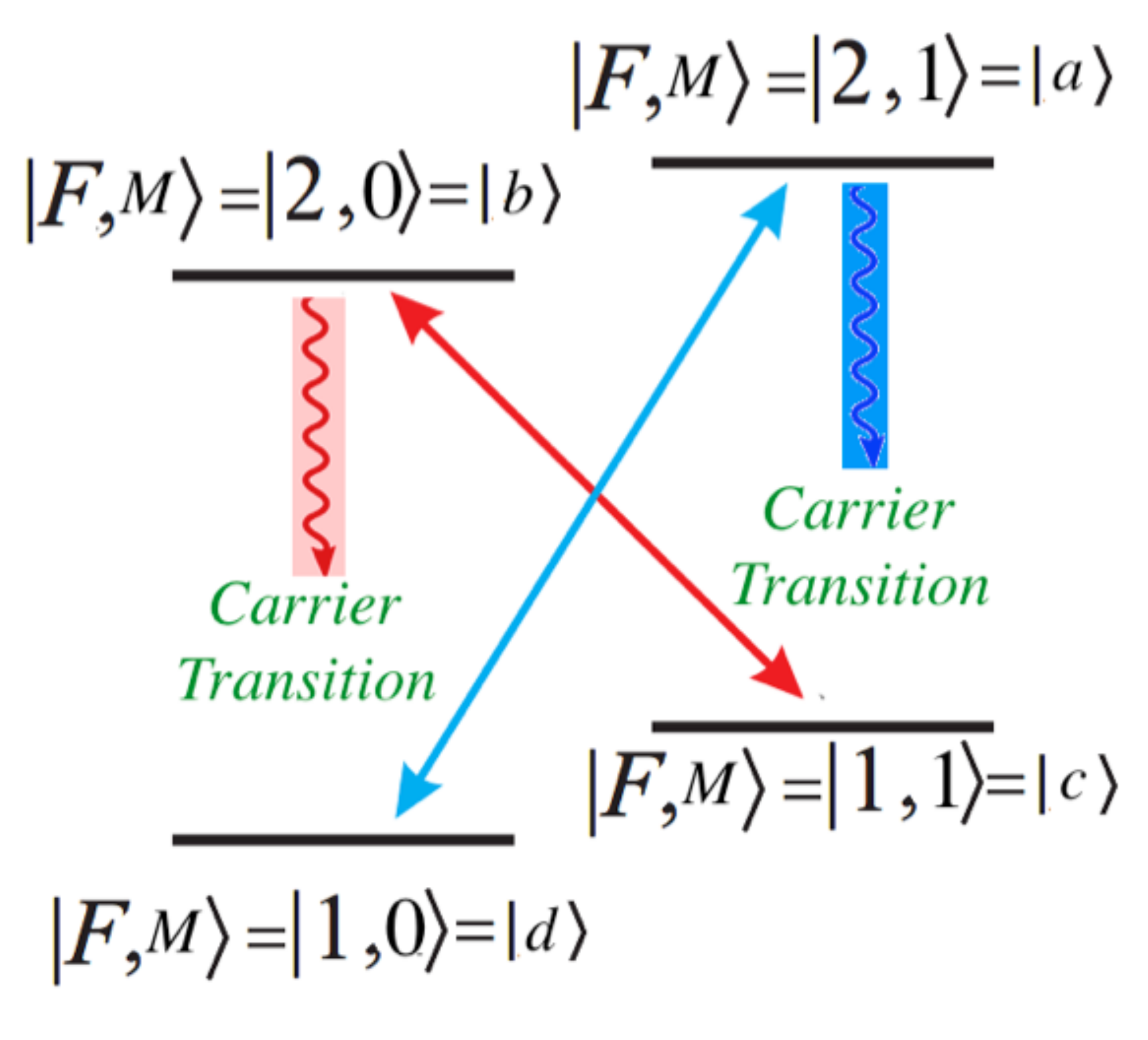}
\vspace{-.5cm}
\caption{Scheme for the hyperfine levels and transitions typical from ground states of the alkali ions. The atomic labels, $\vert F, M\rangle$, assign $F$ to the quantum number for total angular momentum and $M$ to the projection of the angular momentum onto the trap magnetic field direction.}
\label{eqefig:level}
\end{figure}

Considering the rotating wave approximation, $\hat{H}_{RJC}$ includes the Jaynes-Cummings (JC) and anti-Jaynes-Cummings (AJC) interactions through the Hamiltonian
\begin{equation}
\label{eqeA01}
\hat{H}_j ^{(A)JC} = \hbar\, \eta_j \tilde{\Omega}_j \, (\,\hat{\sigma}^{+(-)} a_j e^{+(-)i \phi_{r(b)}} + \hat{\sigma}^{-(+)} a_j^\dagger e^{-(+)i \phi_{r(b)}} \,) + \hbar\, \delta_j \hat{\sigma}_z,
\end{equation}

with $j = x,\,y,\,z$, where the phases $\phi_{r\,(b)}$ describe red(blue)-sideband excitations, $\tilde{\Omega}_j$ are the Rabi frequencies, $\hat{\sigma}^{+ \, (-)}$ are the internal level raising(lowering) operators, $\delta_j$ is an emergent detuning frequency between the external field and the two-level system, and $\eta_j = k \sqrt{\hbar\,/2 \tilde{m} \nu_j}$ is the Lamb-Dicke parameter (where $\tilde{m}$ is the ion mass and $k$ is the wave number of the external field).
From the experimental perspective, this dynamics is reproduced by the hyperfine levels ($2 s^2 \, S_{1/2}$) of alkali ions as depicted in Fig.~\ref{eqefig:level}, where the third interaction is also identified: the carrier interaction given by the Hamiltonian \cite{n012}
\begin{equation}
\label{eqeA03}
\hat{H}_j^C = \hbar\, \Omega_j (\hat{\sigma}^+ e^{i \phi} + \hat{\sigma}^- e^{-i \phi}),
\end{equation}
which accomplishes an excitation of the internal levels and does not change the vibrational state of the ion.

Tthe combination of JC, AJC and carrier interactions can be mapped into a Dirac Hamiltonian including tensor and pseudotensor external fields \cite{n001,n002,n003,n004,n005},
\begin{equation}
\label{DiracHam}
\hat{H}_D = \hat{\beta} \, m c^2 + c\, \hat{\bm{\alpha}}\cdot\bm{p} + \hat{\beta} \hat{\bm{\Sigma}} \cdot \left( \kappa \, \bm{\mathcal{E}} \right) + i \hat{\beta} \hat{\bm{\alpha}} \cdot \left( \mu \, \frac{\bm{\mathcal{E}}}{c} \right),
\end{equation}
from which one firstly notices that the Dirac mass term reads
\begin{equation}
\label{eqeA04}
\hat{\beta} m c^2 \rightarrow 2 \hbar\, \delta( \hat{\sigma}_z^{ad} + \hat{\sigma}_z ^{bc}),
\end{equation}
such that $\hat{\sigma}_z^{ad} + \hat{\sigma}_z ^{bc} \equiv \hat{\beta}$, where the upper indices denote the involved internal levels \cite{n003,n004,n005,nossopaper}.
In the same way, for example, the $p_x$ term of Dirac equation is set by $\phi_r = - \pi/2$ and $\phi_b = \pi/2$ into Eqs.~(\ref{eqeA01}) and, by taking into accont similar choices for other phases, the kinetic term reads
\begin{equation}
\label{eqeA05}
c \, \hat{\bm{\alpha}}\cdot \bm{p} \rightarrow 2 \eta \Delta_x \tilde{\Omega}(\hat{\sigma}_x^{ad} + \hat{\sigma}_x^{bc}) p_x + 2 \eta \Delta_y \tilde{\Omega}(\hat{\sigma}_y^{ad} - \hat{\sigma}_y^{bc})p_y + 2 \eta \Delta_z \tilde{\Omega}(\hat{\sigma}_x^{ac} - \hat{\sigma}_x^{bd}) p_z,
\end{equation}
with $p_j \rightarrow \frac{i \hbar\,}{2 \Delta_j} \, (\,a_j ^\dagger - a_j \,)$,
where $\Delta_j = \sqrt{\hbar\,/ 2 \tilde{m} \nu_j}$ is the delocalizaton width of the ground state wave function and the space homogeneity of the free Dirac equation requires that $\Delta_j = \Delta$.

In addition, the external fields $i \hat{\beta} \hat{\bm{\Sigma}} \cdot \left( \kappa \, \bm{\mathcal{E}} \right)$ and $i \hat{\beta} \hat{\bm{\alpha}} \cdot \left( \mu \, \frac{\bm{\mathcal{E}}}{c} \right)$, which respectively describe tensor and pseudotensor potentials, are mapped in terms of the carrier interactions (\ref{eqeA03}) with frequencies $\Omega_j^{(1)}$ and $\Omega_j^{(2)}$:
\begin{subequations}
\label{eqeA06}
\begin{equation}
\hat{\beta} \hat{\bm{\Sigma}} \cdot \left( \kappa \, \bm{\mathcal{E}} \right) \rightarrow 2 \hbar\, \Omega_x ^{(1)} \, (\, \hat{\sigma}_x^{ab} - \hat{\sigma}_x^{cd}\,) \, + \, 2 \hbar\, \Omega_y ^{(1)} \,(\,\hat{\sigma}_y^{ab} - \hat{\sigma}_y^{cd} \,)\, + \, 2 \hbar\, \Omega_z ^{(1)}\, (\, \hat{\sigma}_z^{ab} - \hat{\sigma}_z^{cd} \,),
\end{equation}
\begin{equation}
i \hat{\beta} \hat{\bm{\alpha}} \cdot \left( \mu \, \frac{\bm{\mathcal{E}}}{c} \right) \rightarrow 2 \hbar\, \Omega_x ^{(2)}\, (\, -\hat{\sigma}_y^{ad} - \hat{\sigma}_y^{bc}\,) \, + \, 2 \hbar\, \Omega_y ^{(2)}\,(\, \hat{\sigma}_x^{bc} - \hat{\sigma}_x^{ad}\,) \, + \,2 \hbar\, \Omega_z ^{(2)} \,(\, \hat{\sigma}_y^{bd} - \hat{\sigma}_y^{ac} \,).
\end{equation}
\label{eqeA06B}
\end{subequations}

The maps from Eqs.~(\ref{eqeA04}), (\ref{eqeA05}) and (\ref{eqeA06B}) reproduce the Dirac dynamics (\ref{DiracHam}) if the relations between the ionic system observables and the Dirac-like parameters are identified by
\begin{equation}
\frac{\mu\, \mathcal{E}_j}{c} = 2 \hbar\, \Omega_j ^{(2)}, \hspace{1.3 cm} \kappa\, \mathcal{E}_j = 2 \hbar\, \Omega_j ^{(1)},\hspace{1.3 cm}
c = 2 \eta \Delta \tilde{\Omega}, \hspace{1.3 cm} m c^2 = 2 \hbar\, \delta,
\end{equation}
through which a one-to-one correspondence between (\ref{DiracHam}) and the sum of the interactions (\ref{eqeA04}), (\ref{eqeA05}) and (\ref{eqeA06B}) is identified. The eigenstates of the Hamiltonian (\ref{DiracHam}) $\vert \psi_{n,s} \rangle$ (with $n,\,s = 0,\,1$), are thus simulated by a quantum superposition of the internal ionic states,
\begin{equation}
\label{eqeA07}
\vert \psi_{n,s} \rangle \rightarrow M^a_{n,s} \vert a \rangle + M^b_{n,s} \vert b \rangle + M^c_{n,s} \vert c \rangle + M^d_{n,s} \vert d \rangle.
\end{equation}
Given that the ionic states exhibit two internal degrees of freedom, one related to the total angular momentum $F$ of the state and other associated to its projection onto the degeneracy lifting magnetic field ${\bm M}$, it is possible to adopt an assignment between the ionic states and two-{\em qubit} states by the correspondence:
\begin{equation}
\label{eqeA08}
\vert a \rangle \equiv\vert 0 \, 0 \rangle, \hspace{1.5 cm} \vert b \rangle \equiv \vert 0 \, 1 \rangle, \hspace{1.5 cm}
\vert c \rangle \equiv \vert 1 \, 0 \rangle, \hspace{1.5 cm} \vert d \rangle \equiv \vert 1 \, 1 \rangle,
\end{equation}
such that the eigenstates (\ref{eqeA07}) are entangled states.

The eigenstates of the Dirac Hamiltonian, $\vert \psi_{n,s} \rangle$, exhibit an intrinsic \textit{spin-parity} entanglement that, from the mathematical point of view, is supported by the bi-spinor structure of the Dirac equation \cite{Salomon,nossopaper}. Considering an $SU(2)\otimes SU(2)$ mapped correspondence of the \textit{spin-parity} two-qubit assignment with two ionic degrees of freedom, such an intrinsic entanglement reproduce the same entanglement profile of ${\bm F}$ and ${\bm M}$ (ionic) degrees of freedom \cite{nossopaper}. Once the dynamics of the bi-spinor $\vert \psi_{n,s} \rangle$ is recovered, the evolution and the entangling properties of any ionic state can be described.
These correlational properties are driven by the $SU(2)\otimes SU(2)$ structure of the Hamiltonian (\ref{DiracHam}), as one can identify by writing the Hamiltonian $\hat{H}_D$ in terms of two-{\em qubit} operators as
\begin{equation}
\label{qubit}
\hat{H}_D = (\hat{\sigma}_z^{(1)} \otimes \hat{I}_2^{(2)}) \, m c^2 + c\, (\hat{\sigma}_x^{(1)} \otimes \hat{\bm{\sigma}}^{(2)})\cdot\bm{p} + (\hat{\sigma}_z^{(1)} \otimes \hat{\bm{\sigma}}^{(2)}) \cdot \left( \kappa \, \bm{\mathcal{E}} \right) - (\hat{\sigma}_y^{(1)} \otimes \hat{\bm{\sigma}}^{(2)})  \cdot \left( \mu \, \frac{\bm{\mathcal{E}}}{c} \right),
\end{equation}
where the superscripts $1$ refers to the total angular momentum subsystem, associated to the first qubit of the assignment (\ref{eqeA08}), and the superscript $2$ refers to the subsystem related to the projection of the total angular momentum onto the lifting magnetic field, associated to the second qubit of (\ref{eqeA08}).

The engineering of internal ionic levels as entangled structures driven by the correspondence between Dirac-like and the trapped ion interactions is supported by a deeper analysis of a larger class of Poincar\'e invariant Dirac-like interactions \cite{n009,nossopaper}. The Hamiltonian eigenstates $\vert \psi_{n,s} \rangle$ indeed exhibits a naturally entangled structure which can be straightforwardly computed from stationary pure states, $\varrho_{n,s} = \vert \psi_{n,s} \rangle \langle \psi_{n,s} \vert$, with $n,\,\, s = 0 \,, \, 1$ corresponding to the two-{\em qubit} assignment (\ref{eqeA08}). The Dirac Hamiltonian (\ref{disc}) includes both tensor and pseudotensor potentials describing a non-minimal coupling with an external constant electric field, and it exhibits algebraic properties which allows for obtaining its eigenstates. From (\ref{DiracHam}) one has
\begin{equation}
\hat{H}_D ^2 = (p^2 + m^2 + (\kappa^2 + \mu^2)\mathcal{E}^2)\hat{I} + 2 \mathcal{O},
\end{equation}
 with
\begin{equation}
\hat{\mathcal{O}} = \frac{1}{2}\left(\hat{H}_D ^2 - \hat{I} \right) = m\,\kappa\, \hat{\bm{\Sigma}} \cdot \bm{\mathcal{E}} + \mu \, \hat{\beta}\,\hat{\bm{\Sigma}} \cdot \, (\, \bm{p} \times \bm{\mathcal{E}} \,) - i \kappa\,\hat{\beta} \,\hat{\bm{\alpha}} \cdot (\, \bm{p} \times \bm{\mathcal{E}} \,),
\label{MathO}
\end{equation}
and $\hat{I}$ the $4 \times 4$ identity operator. Moreover, (\ref{MathO}) satisfies
\begin{equation}
\mathcal{O}^2 = g_2 \hat{I}
\end{equation}
where $g_2$ is evaluated as
\begin{equation}
g_2 = \frac{1}{16}\mbox{Tr}[(\hat{H}_D^2 - \frac{1}{4} \mbox{Tr}[\hat{H}_D^2])^2] = m^2 \kappa^2 \mathcal{E}^2+ (\mu^2 + \kappa^2)(\bm{p} \times \mathcal{\bm{E}})^2.
\end{equation}
The corresponding Hamiltonian eigenvalues can be obtained through the \textit{ansatz} \cite{n009}
\begin{eqnarray}
\label{varho}
\varrho_{n,s} = \frac{1}{4} \left[ I +\frac{ (-1)^n}{\vert \lambda_{n,s} \vert} \hat{H}_D \right] \left[I + \frac{(-1)^s}{\sqrt{g_2}} \hat{\mathcal{O}} \right],
\end{eqnarray}
which corresponds to stationary pure states of the Liouville equation $[\varrho_{n,s}, \hat{H}_D]=0$, and $\lambda_{n,s}$ is identified as the mean energy of $\varrho_{n,s}$, $\lambda_{n,s} = \mbox{Tr} [\, \hat{H}_D \, \varrho_{n,s}\,]$ (see the Appendix for the step-by-step construction of the \textit{ansatz}). Furthermore, considering from now on a Dirac-like one-dimensional propagation along the $x$ axis, with the electric field lying in the $xy$-plane, such that
$
\bm{\mathcal{E}} = \mathcal{E}(\, \cos{\theta}\,\bm{i} + \sin{\theta} \,\bm{j} \,) \nonumber $, with
$\bm{p} = p \,\bm{i}$,
where $\bm{i},\bm{j},\bm{k}$ define an orthonormal basis, one has $\bm{p} \times \bm{\mathcal{E}} = p \, \mathcal{E} \sin{\theta} \, \bm{k}$.
Since the choice of $\theta\neq 0$ does not qualitatively affect the results, a simplifying scenario with $\theta = \pi/4$ can be considered.
In this case, the expressions for $g_2$ and $\lambda_{n,s}$ reads
\begin{subequations}
\begin{eqnarray}
g_2 &=& \mathcal{E}^2 \left[\, m^2 \kappa^2 + \frac{1}{2}(\mu^2 + \kappa^2) p^2 \, \right], \\
\lambda_{n,s} &=& (-1)^n\bigg[\, p^2 + m^2 + (\kappa^2 + \mu^2)\mathcal{E}^2 + 2 (-1)^s \mathcal{E}\sqrt{m^2 \kappa^2 + \frac{1}{2}(\mu^2 + \kappa^2)\,p^2 \,} \bigg]^ {1/2}.
\end{eqnarray}
\end{subequations}

In order to recover the internal ionic state dynamics, one follows a step-by-step construction connecting the Dirac bi-spinor basis, $\{ \vert \, \psi_{n,s} \, \rangle \}$, to the ionic state basis, $\{ \vert\, i \,\rangle \}$. The temporal evolution of a single internal level can be recovered by using the completeness relation $\displaystyle \sum^1_{n,s = 0} \varrho_{n,s} = \hat{I}$ as to have (for $\vert j (t=0) \rangle \equiv \vert j \rangle$)
\begin{equation}
\label{evolution}
\rho_j (t) = \vert\, j (t)\, \rangle \langle\, j(t) \, \vert
= e^{- i \hat{H}_D t} \vert j \rangle \langle j \vert e^{ i \hat{H}_D t} = \displaystyle \sum^1_{n,s=0}\sum^1_{m,l=0} e^{- i (\lambda_{n,s} - \lambda_{m,l}) t}\, \varrho_{n,s} \, \rho_j (0) \, \varrho_{m,l},
\end{equation}
where $\hat{H}_D \varrho_{n,s} = \lambda_{n,s} \, \varrho_{n,s}$. This time evolution of an internal level presents a typical pattern of quantum oscillation phenomena for a four level system. For instance, a state initially prepared as $\vert j \rangle$ oscillates and can be converted into a generic ionic state $\vert k \rangle \neq \vert j \rangle$. By defining the projector $\hat{P}_k = \vert k \rangle \langle k \vert$, the probability of measuring the trapped ion system in such a configurational state is given by
\begin{equation}
\mathcal{P}_{j \rightarrow k} (t) = \mbox{Tr}[\hat{P}_k \, \rho_j (t)] = \displaystyle \sum_{(n,s)=0,1} \sum_{(m,l)=0,1} \, e^{- i (\lambda_{n,s} - \lambda_{m,l}) t} \, \,\mbox{Tr}[\, \hat{P}_k \, \varrho_{n,s} \, \rho_j (0) \, \varrho_{m,l}\, ].
\end{equation}

The energy levels depicted in Fig.~\ref{eqefig:level} and the {\em qubit} assignment from (\ref{eqeA08}) suggest the identification of two subsystems, $\mathcal{S}_F$ and $\mathcal{S}_{M}$ - the former one related to the total angular momentum quantum number, $\bm F$, and the latter one associated to the projection of the angular momentum onto the direction of the lifting magnetic field, $\bm M$.
Within such a framework, an internal ionic state $\vert j \rangle$ will evolve to a superposition between the four ionic states and shall exhibit quantum entanglement between $\mathcal{S}_F$ and $\mathcal{S}_{M_F}$ - which may be detected even from its departing configuration. Moreover, the entangling properties of any initial ionic state can be recovered, once its density matrix evolution is completely described by means of (\ref{evolution}), which can be related with the average chirality of the state \cite{nossopaper}.

\section{Dynamics of Werner and Cat states under global noise}

Once the time evolution of an ionic state is recovered, it is possible to include noise effects arising from the environment coupling. It is assumed that both qubits of the system are collectively coupled to a single environment, with dynamical evolution described by the Hamiltonian $\hat{H}_{\rm Env}$ \cite{Yu01} given by
\begin{eqnarray}
\hat{H}_{\rm Env} &=&
- \frac{1}{2} \mu \, B(t) \left(\, \hat{\sigma}^{(1)}_z \otimes \hat{I} ^{(2)} +  \hat{I} ^{(1)} \otimes \hat{\sigma}_z ^{(2)} \,\right),
\label{eqe01}
\end{eqnarray}
where the superscripts (1) and (2) have the same meaning as in Eq.~(\ref{qubit}), such that the complete dynamics of the ionic system is driven by $\hat{H} = \hat{H}_{RJC} + \hat{H}_{Env}$. In this model, the function $B(t)$ represents stochastic environmental fluctuations which acts equally on both subsystems, i. e. the subsystems associated to $\bm{F}$ and $\bm{M}$ quantum numbers are affected by same random field $B(t)$ through Zeeman-like interactions driven by the operator $\hat{\sigma}_z$ acting on the corresponding subsystem. In the ionic basis $\{\vert a \rangle, \vert b \rangle, \vert c \rangle, \vert d \rangle\}$ the noise Hamiltonian is written as
\begin{eqnarray}
\hat{H}_{\rm Env} &=&
- \mu \, B(t)\, \mbox{Diag}\{1\,\,\,\,\,0\,\,\,\,\,0\,\,-1\}.
\end{eqnarray}

Usually this noise model arises from a couple with a bosonic environment and generically causes a global dephasing of the qubits and affects its coherence properties \cite{Yu01}. For example, in quantum optics one might prepare a pair of polarized photons traveling along partially overlapping fibers. In this setup, the global noise would arise due to random birefingence and the noise would cause a gradual depolarization of the photons, which migh lead to advantagens when specific quantum information protocols are considered \cite{Opt01}.

In the trapped ion setup considered, the global noise (\ref{eqe01}) is one of the main sources of environment effects \cite{FreqEst, NoiseTrap01, NoiseTrap02, NoiseTrap03, NoiseTrap04}, being generated by random fluctuations of the magnetic field required to lift Zeeman degeneracy of the ions \cite{NoiseTrap02}. It is assumed that the stochastic magnetic field $B(t)$ is classical and chacterized by the Markovian conditions
\begin{eqnarray}
\langle B(t) \rangle &=& 0 \nonumber \\
\langle B(t) B(t^\prime) \rangle &=& \frac{\Gamma}{\mu^2} \delta(t - t^\prime),
\label{markov}
\end{eqnarray}
were $\langle \cdot \rangle$ represent the ensemble average and $\Gamma$ is the phase relaxation due to the collective interaction with $B(t)$.

The time-evolving analysis supported by the Hamiltonian of the full system, $\hat{H}$, can be better performed when it is decoupled into two pieces through the interaction picture prescribed by
\begin{equation}
\tilde{\rho}(t) = e^{i\int^t_0 \hat{H}_{\rm Env}(s)ds} \rho(0)
e^{-i\int^t_0 \hat{H}_{\rm Env}ds},
\label{master01}
\end{equation}
with $\rho(0)$ given by the initial conditions and $\tilde{\rho}(t)$ describing the time-evolved density matrix in the interaction picture, which can be obtained as a solution of the master equation given in terms of the Kraus representation \cite{mc,kra,wk}. The Kraus representation allows the inclusion of noise effects and the subsequent description of the correlations in the state. Considering the one global collective noise described by the Hamiltonian of the classical noisy field from (\ref{eqe01}), the solution for the dynamic evolution can be implemented through an operator-sum representation \cite{Yu01}.
By taking the statistical mean of Eq.~(\ref{master01}) assuming the Makovian conditions (\ref{markov}), as prescribed by \cite{Yu01}, one can express the behavior of $\tilde{\rho}(t)$ in a compact way in terms of
\begin{equation}
\label{globalsoln}
\tilde{\rho}(t)={\mathcal E}_D(\rho(0)) = \sum^{3}_{\mu=1}D_\mu^\dag(t)
\rho(0) D_\mu(t),
\end{equation}
where the Kraus operators describing the
collective interaction are given by \begin{equation} \label{model1}
D_1=\begin{pmatrix}
\gamma & 0 & 0 & 0\\
0& 1 & 0 & 0\\
0 & 0& 1& 0\\
0 & 0& 0 & \gamma
\end{pmatrix},~
D_2=\begin{pmatrix}
\omega_1 & 0 & 0 & 0\\
0& 0 & 0 & 0\\
0 & 0& 0& 0\\
0 & 0& 0 & \omega_2
\end{pmatrix},~
 D_3=\begin{pmatrix}
0 & 0 & 0 & 0\\
0& 0 & 0 & 0\\
0 & 0& 0& 0\\
0 & 0& 0 & \omega_3
\end{pmatrix},
\end{equation}
with $\gamma=e^{-\Gamma t/2}$, $\omega_1=\sqrt{1-e^{- \Gamma \,t}}$,
$\omega_2=-\omega_1 e^{-\Gamma \, t}$, and $\omega_3=\omega_1^2\sqrt{1+e^{-\Gamma \, t}}$. To recover the density matrix in the Schr\"odinger picture, $\rho(t)$, the completeness relation for $\{ \varrho_{n,s} \}$ is recovered as to give
\begin{eqnarray}
\label{eqeD01}
\rho(t) = e^{i \hat{H}_D t} \tilde{\rho}(t) e^{ - i \hat{H}_D t} &=& \displaystyle \sum^1_{n,s = 0} \sum^1_{m,l = 0} e^{ - i (\lambda_{n,s} - \lambda_{m,l}) t} \, \rho_{n,s} \, \tilde{\rho} (t) \, \varrho_{m,l} \nonumber \\
&=& \displaystyle \sum^1_{n,s = 0} \sum^1_{m,l = 0} \sum_{\mu=1}^3 e^{ - i (\lambda_{n,s} - \lambda_{m,l}) t} \, \varrho_{n,s} \,D_\mu^\dag(t)\, \rho(0)\, D_\mu(t) \, \varrho_{m,l},
\end{eqnarray}
from which any observable can be evaluated for a given initial state $\rho(0)$. For instance, the survival probability. i. e. the probability of measure a state in its initial configuration, $\mathcal{P}_{\rho(0)} (t)$ reads
\begin{equation}
\label{survivor}
\mathcal{P}_{\rho_0} (t) = \mbox{Tr} [\rho(0) \rho(t)] = \displaystyle \sum^1_{n,s = 0} \sum^1_{m,l = 0} \sum_{\mu=1}^3 e^{ - i (\lambda_{n,s} - \lambda_{m,l}) t} \, \mbox{Tr}[\rho(0) \varrho_{n,s} \,D_\mu^\dag(t)\, \rho(0)\, D_\mu(t) \, \varrho_{m,l}].
\end{equation}

\begin{figure}
\includegraphics[width = 11.5 cm]{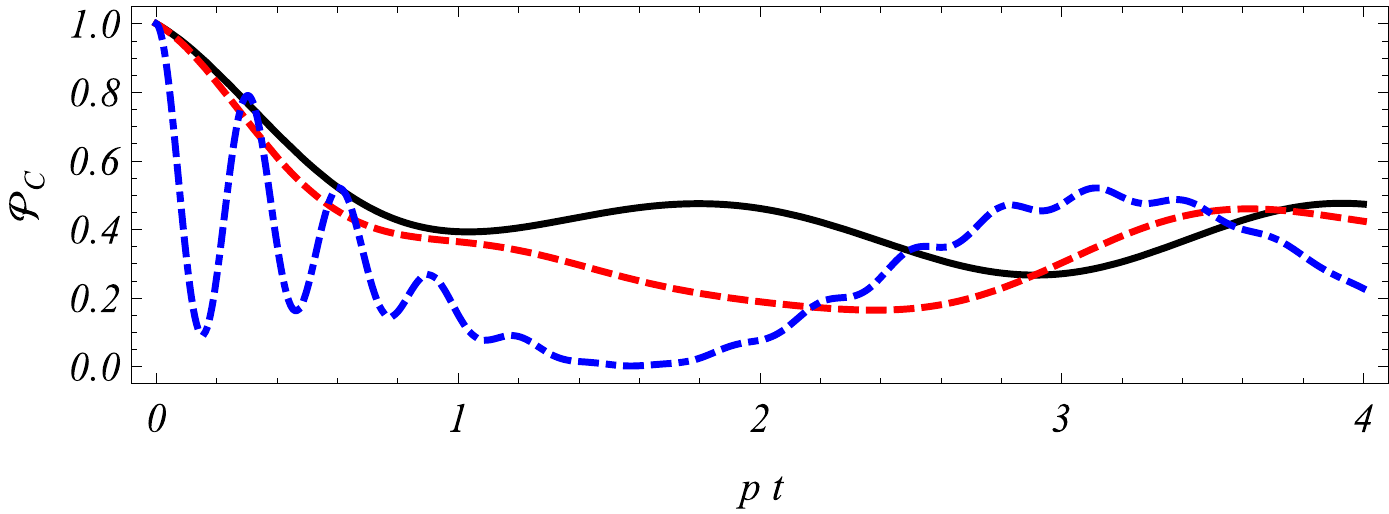}
\includegraphics[width = 11.5 cm]{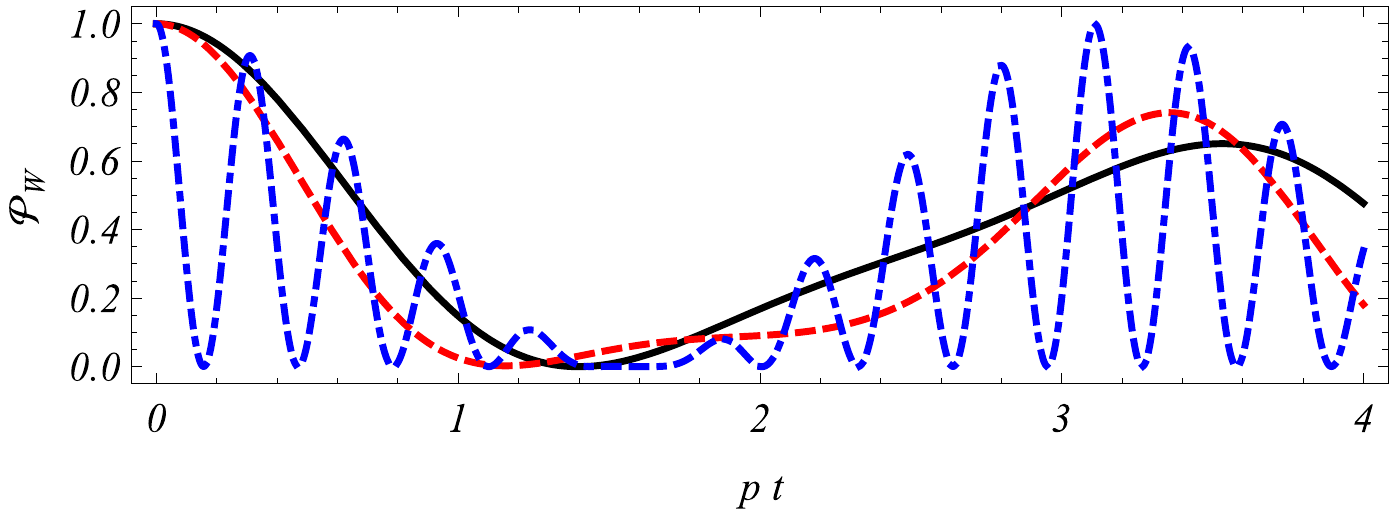}
\caption{Schr\"odinger cat and Werner state survivor probabilities, $\mathcal{P}_{C}$ (lower pannel) and $\mathcal{P}_{W}$ (upper pannel), as function of the dimensionless parameter $p\,t$. The plots are for $\kappa = \mu = 1$, for $\mathcal{E}/p = 1$, $\Gamma/p = 1/2$ and for $m/p = 0$ (solid curves), $1$ (dashed curves), $10$ (dot-dashed curves).
}
\label{eqefig:01}
\end{figure}

From now on, properties under collective noise effects can be investigated for the Schr\"odinger cat state $\rho_C(0) = \vert \psi_C \rangle \langle \psi_C \vert$ and the Werner state $\rho_W (0) = \vert \psi_W \rangle \langle \psi_W \vert$ respectively written in terms of
\begin{eqnarray}
\vert \psi_C \rangle = \frac{\vert a \rangle + \vert d \rangle}{\sqrt{2}}, \hspace{1 cm} \vert \psi_W \rangle = \frac{\vert b \rangle + \vert c \rangle}{\sqrt{2}}.
\end{eqnarray}
The time evolution of these initial states are obtained via Eq.~(\ref{eqeD01}) from which the survivor probabilites $\mathcal{P}_C(t) = \mbox{Tr}[\rho_C(0) \rho_C(t)]$ and $\mathcal{P}_W (t) = \mbox{Tr}[\rho_W(0) \rho_W(t)]$ are recovered by Eq.~(\ref{survivor}), with results depicted in Fig.~\ref{eqefig:01}, as function of the dimensionless parameter $p t$ (in natural units, $p t \sim p t (c/\hbar\,)$).
Each cat state has its initial superposition driven off by the interaction with the collective noisy environment. For $t > 0$, the survivor probability $\mathcal{P}_C (t)$ is always less than $1$. On the other hand, the $\rho_W(t)$ oscillates between $\rho_W (0)$ and an orthogonal superposition to the initial state. Since the ratio between the eigenenergies $\lambda_{n,s}$ does not define a rational number, the Werner state oscillates in time without a defined frequency. The main driver of the overall decoherence effect is the decreasing value of $m/p$ ($\sim \tilde{m}\delta/(k\Omega)$ in the ionic basis), which is more easily evinced for the Werner states. Anyway, decreasing values of $m/p$ drive the suppression of the oscillation pattern for both, cat and Werner states.

The negativity, $\mathcal{N}$, shall be adopted as the entanglement measure \cite{n019,QC01}, from which, the Peres criterion establishes that, in order to have a separable state, all eigenvalues of its partial transpose density matrix must be positive \cite{n019}. The negativity of a state $\rho$ is thus defined by \cite{QC01}
\begin{equation}
\mathcal{N}[\rho] = \vert \vert \, \rho^T_A \, \vert \vert - 1 = \displaystyle \sum_i \vert \mu_i \vert -1,
\end{equation}
where $\vert \vert \, \rho^T_1 \, \vert \vert = \sum_i \vert \mu_i \vert$ stands for the $1$-norm (or trace norm) of the matrix $\rho^T_1$, obtained by the partial transposition of $\rho$ with respect to the subsystem $1$, whose eigenvalues are $\{ \mu_i \}$. Roughly speaking, negativity measures the extent to which the partial transpose fails to be positive.

Separable mixed states can also display more general quantum correlations \cite{QC02}.
The quantum discord, for instance, is a measure of the difference between two expressions for the mutual information which are classically equivalent but different when evaluated in the context of quantum mechanics \cite{QC03}. The calculation of quantum discord involves an optimization process, which is an extremely involved task, even in the two-{\em qubit} case. The geometric discord, $\mathcal{D}$, defined as the minimal Hilbert-Schmidt distance between a given state and the set of zero discord states, has the same qualitative information about the behavior of quantum correlations encoded by the quantum discord, with the advantage of being driven by a simpler computation \cite{QC04}. Given the two-{\em qubit} state $\rho$ in its Fano representation,
\begin{equation}
\rho = \frac{1}{4} \left[ I_4 + (\hat{\bm{\sigma}}^{(1)} \otimes \hat{I}_2^{(2)}) \cdot \bm{a}_1 + (\hat{I}_2^{(1)} \otimes \hat{\bm{\sigma}}^{(2)}) \cdot \bm{a}_2 + \displaystyle \sum_{i,j = 1}^3 t_{ij} (\hat{\sigma}_i^{(1)} \otimes \hat{\sigma}_j^{(2)}) \right],
\end{equation}
the geometric discord is evaluated as \cite{QC04}
\begin{equation}
\mathcal{D}_{1 \, (2)}[\rho] = \frac{1}{4} \left( \, \vert\vert \mbox{\boldmath$a$}_{1 \, (2)}\vert\vert ^2 \, + \vert \vert \, T \, \vert \vert ^2 - k_{max} \right),
\label{disc}
\end{equation}
where $T$ is the matrix with elements $[T]_{ij}= t_{ij}$, $\vert \vert \, T \, \vert \vert ^2 = \mbox{Tr}[\, T \, T^T \,]$ and $k_{max}$ is the largest eigenvalue of the dyadic product ${\mbox{\boldmath$a$}}_{1 \, (2)} {\mbox{\boldmath$a$}}^T_{1 \, (2)} + T \, T^T$. For generalized two-{\em qubit} states, geometric discord and negativity satisfy the inequality
$\mathcal{D}[\rho] \ge \left(\mathcal{N}[\rho] \right)^2$ \cite{QC05}.

\begin{figure}
\includegraphics[width = 11.5 cm]{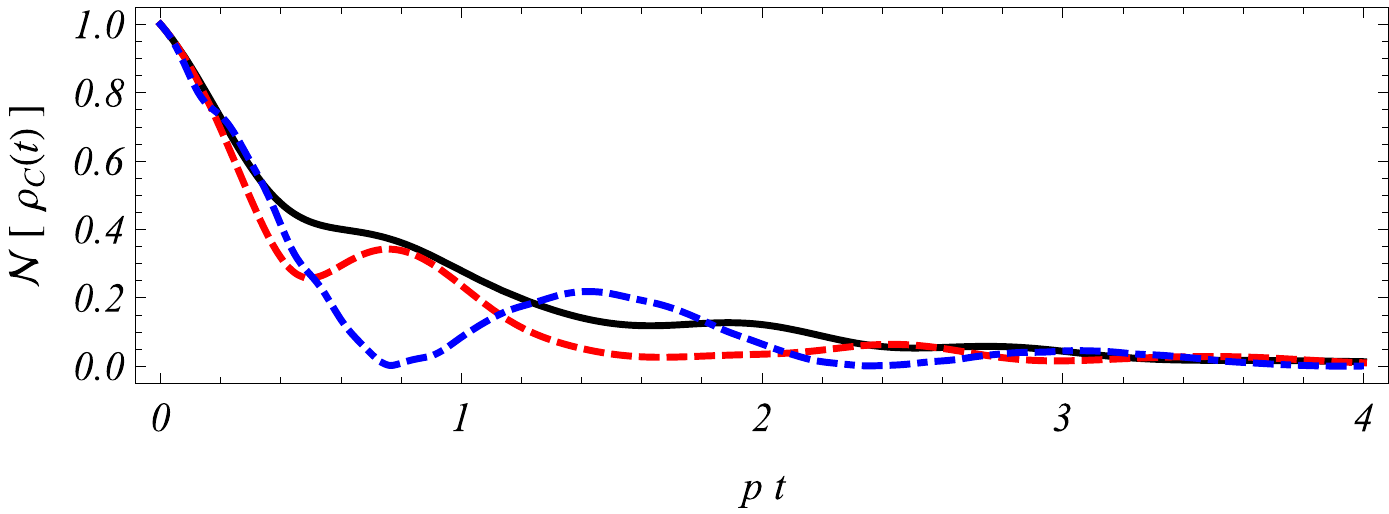}
\includegraphics[width = 11.5 cm]{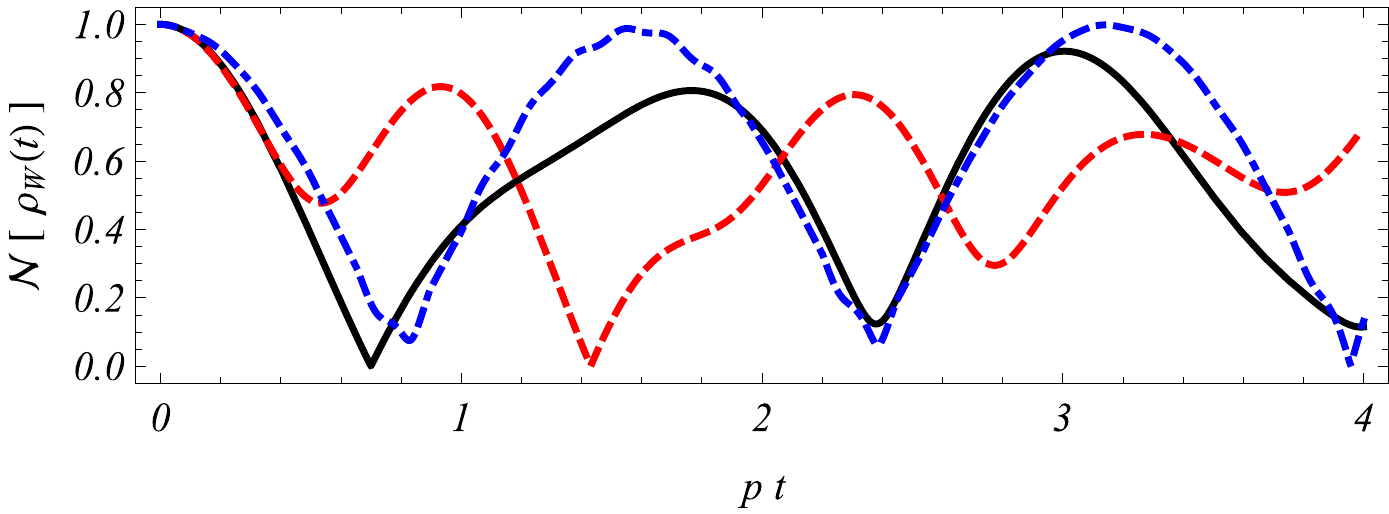}
\caption{Schr\"odinger cat and Werner state negativity, $\mathcal{N}[\rho_C (t)]$ (upper pannel) and $\mathcal{N}[\rho_W(t)]$ (lower pannel), as function of $p\,t$ for the same set of parameters considered in Fig.~\ref{eqefig:01}. Only the Schr\"odinger cat states exhibit a disentanglement profile.}
\label{eqefig:02}
\end{figure}

The negativity for $\rho_W(t)$ and for $\rho_C(t)$ is depicted in Fig.~\ref{eqefig:02}. Werner states are protected against the considered collective noise, as it can be observed by noticing that $ \displaystyle \sum_{\mu = 1}^3 D_{\mu}^\dag (t) \rho_W(0) D_{\mu} (t) = \rho_W(0)$, and therefore $\tilde{\rho}_W (t) = \rho_W (0)$. The Werner state quantum entanglement (lower pannel of Fig.~\ref{eqefig:02}) and the respective survivor probability are not affected by such environment, exhibiting oscillations similar to those reported for the internal levels under an analogous noiseless dynamics \cite{nossopaper}. Moreover, the state is always pure, and therefore the only quantum correlation exhibited is the entanglement.
Different from the Werner states,
as consequence of the interactions with the collective noise, the Schr\"odinger cat states suffer decoherence, and entanglement vanishes for $t \gg \Gamma$, with no entanglement sudden death. Even exhibiting a non-monotone disentanglement profile, with quantum entanglement increasing in some intervals, it disentangles asymptotically and tends to a separable mixed state (upper pannel of Fig.~\ref{eqefig:02}).
\begin{figure}
\includegraphics[width = 11 cm]{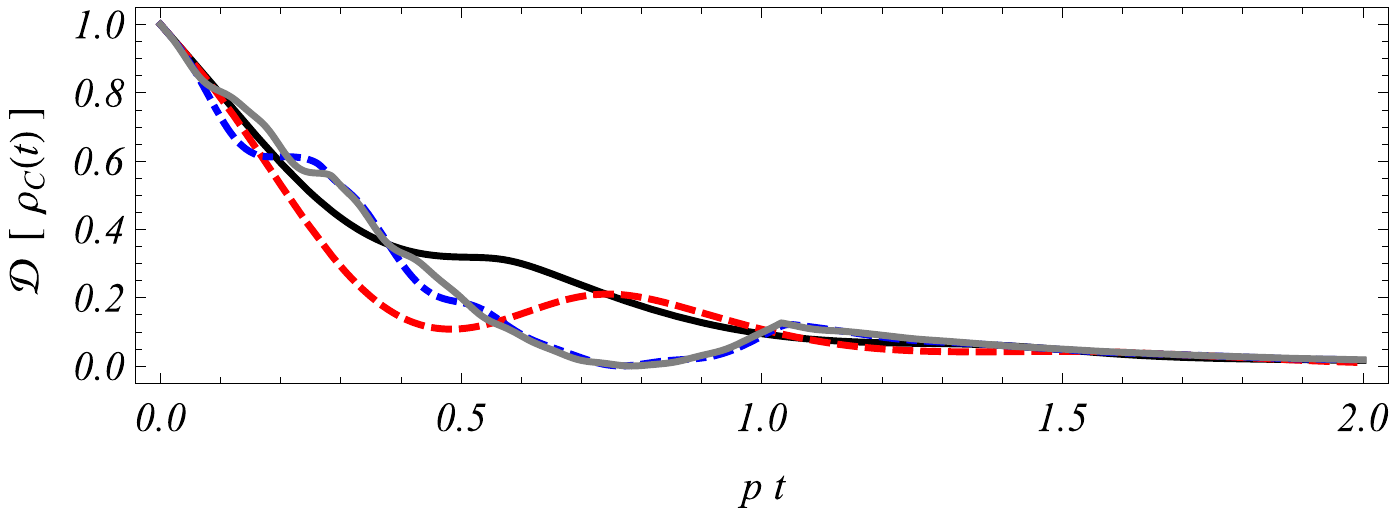}
\includegraphics[width = 11 cm]{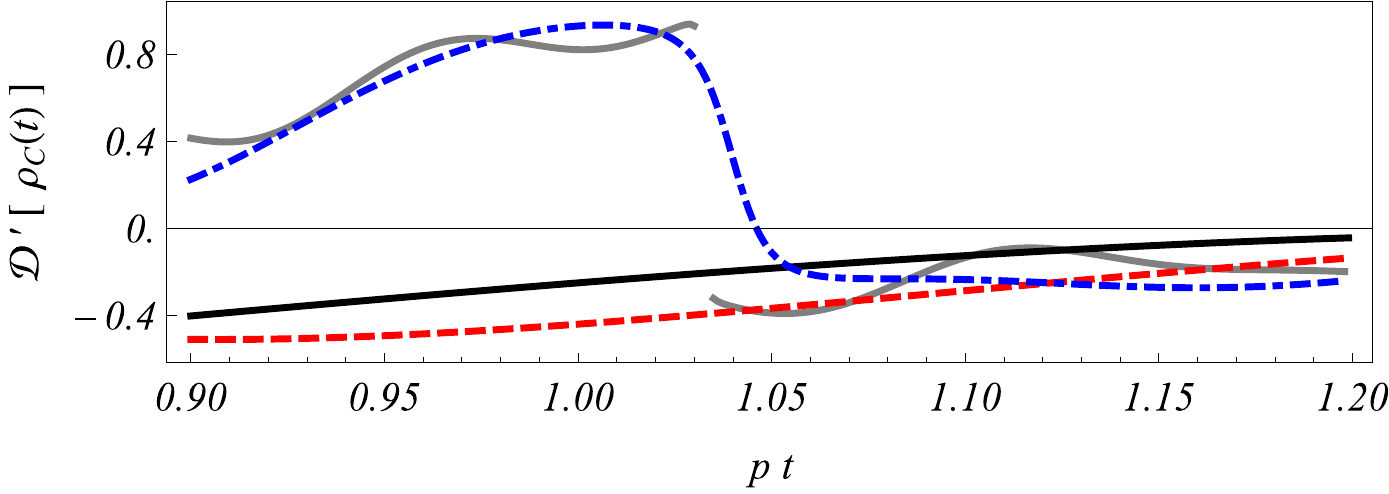}
\caption{Schr\"odinger cat state geometric discord, $\mathcal{D}[\rho_C(t)]$ (upper pannel), and its first derivative $\mathcal{D}^\prime[\rho_C(t)]$ (lower pannel) as function of $p\,t$ for $m/p = 0$ (solid curve), $m/p = 1$ (dashed curve), $m/p= 10$ (dot-dashed curve) and $m/p = 20$ (solid gray curve). All other parameters are in correspondence with those adopted in Figs.~\ref{eqefig:01} and \ref{eqefig:02}. Geometric discord is more robust to environment effects than quantum entanglement, but it also vanishes for $t \gg 1/\Gamma$. For high values of $m/p$, the derivative of geometric discord is discontinuous, indicating a transition between classical and quantum decay regimes of quantum correlations \cite{GeoDisc01,GeoDisc02,GeoDisc04}.}
\label{eqefig:03}
\end{figure}

Finally, the geometric discord for the cat state is depicted in Fig.~\ref{eqefig:03}. Although quantum correlations are more robust to collective noise effects, they also vanish for $t \gg 1/\Gamma$. A striking feature identified along the evolution of geometric discord is the existence of a discontinuity in its first derivative for high values of $m/p$, as it can be noticed in Fig.~\ref{eqefig:03}, for $m/p = 20$ (gray curve, which has a cusp, in correspondence to the lower pannel of Fig.~\ref{eqefig:03}).
Different from quantum entanglement, geometric discord does not abruptly vanish, since the set of zero discord state is a zero measurement set \cite{QC06}.
Meanwhile, the decay of quantum correlations can be driven by different rates, associated either with classical or with quantum decay regimes.
When the transitions between these two regimes are abrupt, quantum discord presents a discontinuity in its first derivative, as for instance, when one consider a bit-phase flip noisy channel \cite{GeoDisc01,GeoDisc02,GeoDisc04}, or even when multi-{\em qubit} states are considered. To our knowledge, is has been reported for the collective noisy channel at a very first time.

The Werner and Cat states can be prepared in an experimental setup by usual techniques. The vibrational ground state of the trapped ion is prepared by laser cooling and the internal ionic configuration can be initialized by optical pumping with a high probability \cite{exp01}, with its fedility limited by the quality of the driven laser polarization \cite{n012}. Furthermore, electron shelving method, based on detecting laser-induced fluorescence on an electric dipole allowed transition, can be used to detect the internal ionic states. Such technique has been used, for example, to measure the simulation of \textit{zitterbewegung} effect in a trapped ion setup, where the position operator $\langle \hat{x} \rangle$ of the relativistic system has been mapped into the internal levels of the ion \cite{n003}.

The experimental measurement of quantum correlations is a involving task. Negativity can not be directly measure, since it is given through the partial transposition, which is an unphysical operation. Nevertheless quantum entanglement can be measure through specific properties of a given state. For pure states and a restricted class of mixed states, entanglement can be measured through the hyperentanglement between copies of the system \cite{hyperent01, hyperent02, hyperent03, hyperent04}. In such setup, one consider a system composed of two copies of the same quantum system, and quantum concurrence (an entanglement measurement) can be measured through a single auxiliary photon measurement, in similar way as perscribed by \cite{hyperent02}. For instance once $\vert \psi_W \rangle$ is not affected by the noise, such protocol can be applied to measure its entanglement. On the other hand, $\vert \psi_C \rangle$ evolves into a mixture, and the protocol can not be applied to detect its entanglement. The measurement of entanglement in this case requires a complete knowledge of the states density matrix, which can be achieved by quantum state tomography, that was implement in different trapped ion setups \cite{tomograph01, tomograph02, tomograph03}. The evaluation of geometric discord also requires, in principle, quantum state tomography, which was measured in a liquid state nuclear magnetic resonance setup and that exhibit a discontinuity in its first derivative \cite{NMR} similar to that depicted in Fig.~\ref{eqefig:03}.

A last remark about the dynamics of the mapped quantum system discussed in this paper is related to the ion's position, as the A(JC) interactions (1) couple the motional state of the ion with the internal levels. Although the localization properties of the state have not been evaluated through our approach, to simulate relativistic effects such as the {\em zitterbewergung} and position dependent potentials, the motional degree of freedom of the trapped ion usually considered as to map the localization properties of the Dirac bi-spinor wave function \cite{n004,n006,n001,n002}. From the experimental perspective, there are several protocols to control and measure the motional state of the trapped ion with high precision \cite{intronoise00,motionalheating01}, nevertheless motional heating can occur, for instance, due to electric field noise near the metallic surface of the trapping \cite{motionalheating02}. Motional heating leads to a decoherence of the motional state \cite{motionalheating02A,motionalheating02B,motionalheating03} accompanied by undesired transitions among the internal levels of the ion \cite{intronoise00,motionalheating03}. Essentially, it induces some transitions among levels not involved in the simulation protocol, modifying the probability of the state to be in a specific energy level \cite{motionalheating04} and changing the quantum superposition that describes the state, affecting its correlational content. In this case, more degrees of freedom of the trapped ion would be excited, and more internal levels would be relevant in modifying the bipartite structure (or maybe changing it into a multipartite one) which has been investigated here.
That is a point which deserve a more careful analysis in the scope of quantifying some additional quantum correlation properties.

\section{Conclusions}

To conclude, the issues investigated throughout this letter were concerned with the inclusion of collective noise effects, described by a stochastic magnetic field, acting on four ionic levels, in the context of a Dirac-like dynamics simulated trapped ion systems, for which quantum correlation properties were quantified in terms of negativity and geometric discord.
Preliminarily, the noiseless dynamics of ionic states was recovered through the correspondence between JC, AJC and carrier interactions, and the Dirac equation structure with the inclusion of tensor and pseudotensor external fields.
The collective noise was introduced by means of the Kraus representation formalism, such that survivor probabilities and quantum entanglement of states prepared as internal levels of the trapped ion were then investigated.
The quantum correlational properties exhibited by Werner and Schr\"odinger cat states formed by superpositions between two internal ionic levels were finally obtained.
Because the states are not eigenstates of the total Hamiltonian of the system, survivor probabilities oscillates in time: the cat states had their initial configuration driven off and the Werner states reproduced an oscillating behavior which generally does not exhibit an identifiable periodicity.
Generically, the Werner states seems to be protected against collective noise effects, and did not exhibit entanglement loss due to the interaction with the environment. The entanglement oscillations in such states are due to the typical structure of the noiseless evolution under the JC, AJC and carrier interactions \cite{nossopaper}. On the other hand, the Schr\"odinger cat states asymptotically disentangle due to the influence of the collective noise. There is no occurrence of entanglement sudden death, and the entanglement loss is not strictly decreasing i. e. there are periods in which entanglement actually increases, which is also a reflection of the noiseless evolution structure of such states. Besides the entanglement loss, quantum correlations quantified by the geometric discord also asymptotically vanish. The geometric discord exhibits a discontinuity along the first derivative for the equivalent to non-relativistic states ($m/p \rightarrow \infty$), which indicates a sudden transition between classical and quantum regimes of decoherence.

To end up, the single trapped ion platforms have worked to implement more complex quantum simulations \cite{Gerritsma}, where the only observable that can be experimentally measured by fluorescence techniques is $\hat{\sigma}_z$ (cf. Eqs. (\ref{eqeA01}) and (\ref{eqe01})).
Since laser pulses can be used	to project other observables onto $\hat{\sigma}_z$ \cite{Gerritsma}, the engendering of a novel state-dependent displacement operation can be investigated in the scope of monitoring time evolution under noise effects discussed in this letter, as identify manipulable mechanisms of disentanglement and even the possibility of some transition between classical and quantum regimes described by this framework.

{\em Acknowledgments - The work of AEB is supported by the Brazilian Agencies FAPESP (grant 15/05903-4) and CNPq (grant 300809/2013-1). The work of VASVB is supported by the Brazilian Agency CNPq (grant 140900/2014-4).}

\pagebreak

\section*{Appendix - Dirac Hamiltonian eigenstates and eigenvalues}

To recover the eigenvalues and the corresponding eigenstates of a specific class of Dirac-like Hamiltonians, one might use a suitable \textit{ansatz} following the procedure derived in Ref.~\cite{n009}. The generalized Hamiltonian, which includes all possible external field global (non-local) couplings, reads
\begin{equation}
\label{appendix01}
\hat{H}_G =  \hat{\beta} m + \hat{\bm{\alpha}}\cdot \bm{\mathcal{P}} + i \hat{\beta} \hat{\gamma}_5 \nu -\hat{\gamma}_5 q  + \hat{\gamma}_5 \hat{\bm{\alpha}} \cdot \bm{\mathcal{W}} + i \kappa_a \hat{\beta} \hat{\bm{\alpha}} \cdot \bm{\mathcal{B}} - \mu_a \hat{\beta} \hat{\bm{\Sigma}} \cdot \bm{\mathcal{B}},
\end{equation}
were $$\hat{\gamma}_5 = \begin{pmatrix}
0 & \hat{\sigma}_x \\
\hat{\sigma}_x & 0 \end{pmatrix}.$$
The external fields included into this dynamics are classified according to their symmetry properties under Poincar\`{e} transformations. Into the above generalized Hamiltonian, one has: the free particle term, $\hat{\beta} m + \hat{\bm{\alpha}}\cdot \bm{\mathcal{P}}$; the pseudoscalar interaction contribution, $ i \hat{\beta} \hat{\gamma}_5 \nu$; the pseudovector interaction contribution, $-\hat{\gamma}_5 q  + \hat{\gamma}_5 \hat{\bm{\alpha}} \cdot \bm{\mathcal{W}}$; and tensor and pseudotensor interaction contributions summarized by $ i \kappa_a \hat{\beta} \hat{\bm{\alpha}} \cdot \bm{\mathcal{B}} - \mu_a \hat{\beta} \hat{\bm{\Sigma}} \cdot \bm{\mathcal{B}}$. For instance, the Hamiltonian (\ref{DiracHam}) is a special case of (\ref{appendix01}) for $\nu = q  = 0$, $\bm{\mathcal{W}} = \bm{0}$, $\kappa_a \bm{\mathcal{B}} \rightarrow \kappa \mathcal{\bm{E}}$ and $\mu_a \bm{\mathcal{B}} \rightarrow \mu \mathcal{\bm{E}}/c$.

Instead of solving directly the Dirac equation with an arbitrary potential matrix, another approach may be adopt as to derive pure state density matrices associated to the eigenstates of (\ref{appendix01}). For $\hat{H}_G$, one has $\mbox{Tr}[\hat{H}_G] = 0$ and the following relations are observed
\begin{eqnarray}
\label{appendix02}
\hat{H}_G^2 &=& c_1 \hat{I}_4 + 2 \mathcal{O}, \nonumber \\
\frac{(\hat{H}_G^2 - c_1 \hat{I}_4)^2}{4} &=& \mathcal{O}^2 = c_2 \hat{I}_4 + 2 [(\nu \kappa_a - m \mu_a)(\bm{\mathcal{W}} \cdot \bm{\mathcal{B}}) - q (\bm{\mathcal{P}} \cdot \bm{\mathcal{W}})] \hat{H}_G,
\end{eqnarray}
were
\begin{eqnarray}
c_1 &=& \frac{1}{4} \mbox{Tr}[ \hat{H}_G^2], \nonumber \\
c_2 &=& \frac{1}{16} \mbox{Tr}\left[ \left(\hat{H}_G^2 - \frac{1}{4} \mbox{Tr}[ \hat{H}_G^2] \right)^2 \right].
\end{eqnarray}
The traceless operator $\hat{\mathcal{O}}$ is given by
\begin{eqnarray}
\mathcal{O} &=& \hat{\bm{\Sigma}} \cdot [\, (\nu \kappa_a - m \mu_a) \bm{\mathcal{B}} - q \bm{\mathcal{P}} \,] + \hat{\beta} \hat{\bm{\Sigma}} \cdot[\, m \bm{\mathcal{W}} + \kappa_a \bm{\omega}_\mathcal{B}] + i \hat{\beta} \hat{\bm{\alpha}} \cdot[\, \nu \bm{\mathcal{W}} + \mu_a \bm{\omega}_\mathcal{B}] \nonumber \\
&-& q \hat{\bm{\alpha}} \cdot \bm{\mathcal{W}} + (\bm{\mathcal{P}} \cdot \bm{\mathcal{W}}) \hat{\gamma}_5 - \mu_a (\bm{\mathcal{W}} \cdot \bm{\mathcal{B}}) \hat{\beta} + i \kappa_a (\bm{\mathcal{W}} \cdot \bm{\mathcal{B}}) \hat{\beta} \hat{\gamma}_5,
\end{eqnarray}
with $\bm{\omega}_\mathcal{B} = \bm{\mathcal{P}} \times \bm{\mathcal{B}}$. The constants $c_1$ and $c_2$ are evaluated as
\begin{eqnarray}
\label{appendix03}
c_1 &=& \mathcal{P}^2 + m^2 + \nu^2 + q^2 + \mathcal{W}^2 + (\kappa_a^2 + \mu_a^2) \mathcal{B}^2, \nonumber \\
c_2 &=& [\, (\nu \kappa_a - m \mu_a) \bm{\mathcal{B}} - q \bm{\mathcal{P}} \,]^2 +[\, m \bm{\mathcal{W}} + \kappa_a \bm{\omega}_\mathcal{B}]^2 + [\, \nu \bm{\mathcal{W}} + \mu_a \bm{\omega}_\mathcal{B}]^2 \nonumber \\
&+&  q^2 \mathcal{W}^2 + (\bm{\mathcal{P}} \cdot \bm{\mathcal{W}})^2 + (\kappa_a + \mu_a)^2 (\bm{\mathcal{W}} \cdot \bm{\mathcal{B}})^2,
\end{eqnarray}
and $\mbox{Tr}[\hat{\mathcal{O}} \hat{H}_G] = 0$.

The eigenvalues $\lambda_{n,s}$ and the eigenstates $\vert \psi_{n,s} \rangle$ of $\hat{H}_G$ are given by the equation $\hat{H}_G \vert \psi_{n,s} \rangle = \lambda_{n,s} \vert \psi_{n,s} \rangle$ or, in terms of the density matrix $\varrho_{n,s} = \vert \psi_{n,s} \rangle \langle \psi_{n,s} \vert$ $$\hat{H}_{G} \varrho_{n,s} = \lambda_{n,s} \varrho_{n,s}.$$ Moreover, the density matrix associated to an eigenstate of $\hat{H}_G$ is a stationary solution of the Liouville equation $[ \hat{H}_G, \varrho_{n,s}] = 0$. From (\ref{appendix02}) one has
\begin{equation}
\label{appendix05}
\hat{\mathcal{O}}^2 \varrho_{n,s} = \frac{1}{4}(\lambda_{n,s}^2 - c_1)^2 \varrho_{n,s},
\end{equation}
and, if $\hat{\mathcal{O}}^2 = c_2 \hat{I}$, then by taking the trace of (\ref{appendix05}) the eigenvalues are recovered as:
\begin{equation}
\lambda_{n,s} = (-1)^n \sqrt{c_1 + 2 (-1)^s \sqrt{c_2}},
\end{equation}
were it was used that  $\mbox{Tr}[\varrho_{n,s}] = 1$. The condition $\hat{\mathcal{O}}^2 = c_2 \hat{I}$ can be accomplished by an adequate choice of parameters, such as by the identification of some relative orientations of the vectors $\bm{\mathcal{P}}$, $\bm{\mathcal{W}}$ and $\bm{\mathcal{B}}$.

To recover the density matrices associated to the eigenstates of $\hat{H}_G$, one firstly notices that the condition $[ \hat{H}_G, \varrho_{n,s}] = 0$, and the fact that for $c_2 \neq 0$ the eigenvalues are non-degenerate, implie that $\varrho_{n,s} = \displaystyle \sum_{i = 0} ^N \xi_i \hat{H}^{i}$ , where $\xi_i$ are real numbers. Due to the imposed condition $\hat{\mathcal{O}}^2 = c_2 \hat{I}$, the potencies of the Hamiltonian satisfy:
\begin{eqnarray}
\hat{H}_G^3 &=& c_1 \hat{H}_G + 2 \hat{\mathcal{O}} \hat{H}_G \nonumber  \\
\hat{H}_G^4 &=& (4 c_2 - c_1^2) \hat{I} + 2 c_1 \hat{H}_G^2 \nonumber  \\
\ldots
\end{eqnarray}
and thus, the density matrix is a 3rd degree polynomial of $\hat{H}_G$:
\begin{equation}
\label{appendix06}
\varrho_{n,s} = \xi_0 \hat{I} + \xi_1 \hat{H}_G + \xi_2 \hat{H}_G^2 + \xi_3 \hat{H}_G^3.
\end{equation}
By evaluating $\mbox{Tr}[\hat{H}_G^i \varrho_{n,s}] = \lambda_{n,s}^i$, for $i = 0, 1, 2, 3$, and using that $\mbox{Tr}[\hat{H}_G^{2 i + 1}] = 0$, one has the following system of equations satisfied by  $\{\xi_0, \xi_1, \xi_2, \xi_3 \}$:
\begin{eqnarray}
\begin{cases}
4 \xi_0 + 4 c_1 \xi_2 = 1 \\
4 c_1 \xi_0 + 4 (4 c_2 + c_1^2) \xi_2 = \lambda_{n,s}^2 \\
4 c_1 \xi_1 + 4 (4 c_2 + c_1^2)  \xi_3 = \lambda_{n,s} \\
4 (4 c_2 + c_1^2)  \xi_1 +  4 [c_1 (4 c_2 - c_1^2) + 2 c_1 (4 c_2 + c_1^2)]\xi_3 = \lambda_{n,s}^3
\end{cases},
\end{eqnarray}
for which the solutions are
\begin{eqnarray}
\xi_0 &=& \frac{1}{4} \left[1 - \frac{c_1 (-1)^s }{2  \sqrt{c_2}} \right], \hspace{1 cm} \xi_1 = \frac{1}{4}\left[\frac{(-1)^n}{\vert \lambda_{n,s} \vert} - \frac{c_1(-1)^{s+ n}}{2 (-1)^s \sqrt{c_2} \, \vert \lambda_{n,s} \vert} \right], \nonumber \\
\xi_2 &=& \frac{ (-1)^s}{8 \sqrt{c_2}}, \hspace{2.8 cm} \xi_3 = \frac{ (-1)^{s+n}}{8\sqrt{c_2} \, \vert \lambda_{n,s} \vert}.
\end{eqnarray}
By substituting these solutions on (\ref{appendix06}), the properties (\ref{appendix02})--(\ref{appendix03}) can be used to write $\varrho_{n,s}$ as the simplified expression
\begin{equation}
\label{appendix04}
\varrho_{n,s} = \frac{1}{4}\left(\hat{I}_4 + \frac{(-1)^s}{\sqrt{c_2}} \hat{\mathcal{O}} \right) \left( \hat{I}_4 + \frac{(-1)^n}{\vert \lambda_{n,s} \vert} \hat{H}_G \right),
\end{equation}
with $n,s = \{1,2\}$. If $\hat{\mathcal{O}} = 0$, then $$\mbox{Tr}[\varrho_{n,s}^2] = \frac{1}{4}\left(1 + \frac{c_1}{\lambda_{n,s}^2}\right),$$ and the \textit{ansatz} is a mixed state.

This procedure for obtaining eigenstates of a Dirac-like Hamiltonian have been used to derive correlation properties of Dirac-like systems \cite{nossopaper} involving several combinations of external fields, as well as the correlational properties driven by them \cite{n009}.

\end{document}